**Tricks of the light: the remarkable power of laser tweezers to dissect complex biological questions**

Mark C Leake[1]

[1] Departments of Physics and Biology, University of York, York YO10 5DD, UK.

Email correspondence to: mark.leake@york.ac.uk

*On 2 October 2018* Göran Hansson, Secretary General of the Royal Swedish Academy of Sciences, announced that *the Nobel Prize in Physics would be jointly awarded to* Arthur Ashkin, Gérard Mourou and Donna Strickland, for their "groundbreaking inventions in the field of laser physics". Strickland and Mourou shared one half of the prize for their pioneering work in generating high-intensity, ultra-short optical pulses. The recipient of the other half was Arthur Ashkin for his seminal work leading to the development of optical tweezers, also referred to as 'optical traps' or 'laser tweezers', and their applications to an enormous range of biological systems. As discussed below, laser tweezers are a remarkable class of optical force transduction tools which have had a profound effect in enabling several complex biological questions to be addressed impenetrable using other existing technologies.

The ability to trap particles using laser radiation pressure was demonstrated by Arthur Ashkin while he worked in the Bell Labs almost half a century ago, using a relatively unstable 1D optical trap consisting of two juxtaposed laser beams whose photon flux resulted in equal and opposite forces on a micron sized glass bead (*Ashkin, 1970*). The modern form of the now most commonly employed 3D optical trap design (which Ashkin denoted as a 'single-beam gradient force trap', *Ashkin, 1986*), results in a net optical force on a refractile, dielectric particle which has a higher refractive index than the surrounding medium, directed approximately towards the intensity maximum of a focused laser beam. The key physics principles that account for the operation of laser tweezers are a testament to the successes of the wave-particle duality model of light; as predicted by quantum mechanics, photons of light carry linear momentum $p$ given by the de Broglie relation which relates this particle property of momentum to the light's wavelength through $p=E/c=hv/c=h/\lambda$, for a wave of energy $E$, frequency $v$ and wavelength $\lambda$ where $c$ is the speed of light and $h$ Plank's constant. This momentum results in radiation pressure if photons are scattered from an object, but also if refraction occurs at the point of a photon emerging from an optically transparent particle there is a deviation in beam direction, and thus a change in the momentum vector resulting in an equal and opposite force on the particle, in accordance with Newton's 3$^{rd}$ law.

Standard laser tweezers have a Gaussian intensity profile in the vicinity of the laser focus which results in a net force that literally traps the particle in a potential well whose width is limited by the diffraction of light, resulting in a trap diameter in the focal plane of a light microscope of roughly the wavelength of the laser light. Many modern day laser tweezers use near infrared laser sources of wavelengths around a micron, ideal for trapping commercially available microscopic spheres composed of glass or latex/polystyrene. Although single biomolecules themselves cannot be easily optically trapped with any great efficiency (some early laser tweezers experiments toyed with rather imprecise manipulation of chromosomes and other cellular structures), they can be manipulated via such micron sized optically trapped spheres. Such microspheres can be controllably coated in a range of chemicals that can enable specific biological molecules to stick to their surface. These coated microspheres can then be used to probe the forces involved in the activities of these biological molecules, either in isolation or coupled to interactions with other molecules, detected optically through measuring the small displacements in the microsphere position relative to the centre of the laser tweezers force field. Laser tweezers in effect act as a microscopic optical force transduction tool *(Leake, 2016)* which can tease out the tiny piconewton level forces generated in changes in single molecule conformations in real time, with an exceptional sensitivity capable of detecting molecular displacements down to the sub-nanometre length scale of individual chemical bonds.

These optical force transduction devices have since been applied to very diverse studies in the area of single molecule biology (*Lenn, 2012*), including biomechanical polymers such as nucleic acids and a range of filamentous proteins, as well as intense research of molecular machines, such as those involved in the generation of force in muscle contraction, molecular trafficking inside cells, and those used to enable cells to be motile (*Miller, 2018*). With high stiffness laser tweezers a single biological molecule can be tethered between a trapped microsphere and a microscope coverslip or another independently trapped microsphere, and controllably stretched and relaxed to explore its viscoelastic properties at a single-molecule level, which has given us invaluable insights into the mechanics of molecules such as DNA and RNA and a range of long proteins such as those of muscle tissue (*Leake, 2003*; *Leake, 2004*). Similarly, laser tweezers have been used in a so-called 'dumbbell assay' originally designed to study 'motor protein' interactions between the muscle proteins myosin and actin (*Finer, 1994*), but since utilized to study several different 'linear' motor proteins (motor protein which operate on a linear molecular track) including kinesin, dynein and DNA-protein complexes. Here, the appropriate molecular track can be tethered between two optically-trapped microspheres and is then lowered onto a third surface-bound microsphere coated in motor protein molecules resulting in stochastic interactions during 'power-stokes' of the motor protein against the track, which may be measured by monitoring the displacement fluctuations of the trapped microspheres.  Very low stiffness laser tweezers can also be used not to manipulate biomolecules but more as a very high precision positional displacement detector through a process of laser interferometry as a trapped microsphere moves relative to the laser focus. By using the microsphere as a probe attached to the flagella of bacteria it has been possible to monitor individual step-like power-strokes of 'rotary' motor proteins (for which the track is in effect a circle) in the same way as linear motor proteins (*Sowa, 2005*). These most basic of laser tweezers devices have given us huge insights into the *Physics of Life* at the single-molecule level (*Leake, 2013*).

The range of capabilities of laser tweezers have expanded significantly over the past *ca.* decade. The refractive index of the inside of cells is in general heterogeneous, with a mean which is marginally higher than the water-based solution of the external environment. This combined with the fact that cells have a mechanical compliance results in an optical stretching effect in these optical fibre based laser tweezers devices, which has been used to investigate mechanical differences between normal human cells and those which have a marginally different stiffness due to being in a diseased state, such as those involved in cancer (*Gück, 2005*).  Also, laser tweezers have been adapted to measure not only force but also molecular *torque*. In standard Gaussian profile laser tweezers there is zero net angular momentum about the optic axis due to the symmetry of the microsphere and the intensity profile of the trap, however rotational asymmetry can be introduced to manipulate and measure angular momentum. For example, two optically trapped microspheres can be fused together to generate a provides a wrench-like 'optical spanner' effect, which has been used for studying the F1-ATPase enzyme (*Pilizota, 2007*), one of the rotary motor proteins which, when coupled to another rotary motor protein called Fo, generates molecules of the 'universal biological fuel' ATP. Another method utilizes the angular momentum properties of light itself. *Laguerre-Gaussian* beams can be generated from higher-order laser modes above the normal symmetrical Gaussian profile used in standard laser tweezers, by either optimizing for higher-order lasing oscillation modes from the laser head or by applying phase modulation optics to the beam path. Combining such asymmetrical beam profiles with the use of circularly polarized light can induce controllable rotation in birefringent particles which can be used to generate torque to study the underlying mechanisms of action of rotary motor, for example in the interactions of certain proteins with DNA (*Forth, 2011*).

The very high detection sensitivity, spatial precision and high time resolution of laser tweezers predispose them to a broad range of applications for studying the mechanics of biological processes at a single-molecule level. The use of focused laser light in generating an optical trap also facilitates invaluable integration into existing light microscopes, thereby enabling complementary imaging observations not only to visualize trapped microspheres but also to visually detect a range of associated biological structures that may interact with the trapped microsphere, for example utilising various forms of fluorescence microscopy that can illuminate specific features of interest. This use of correlative technology offers enormous potential at combining the breath-taking power of molecular manipulation of laser tweezers with the exceptional imaging precision of the emerging suite of phenomenal super-resolution light microscopy tools. The transformative insights into molecular biomechanics, fundamental to basic biological processes, thanks to the power of laser tweezers, is Ashkin's tremendous legacy to cutting-edge interdisciplinary science at the exciting interfaces between the life and physical sciences.